\documentclass[iop,apj]{emulateapj}
\pdfoutput=1
\usepackage[colorlinks,urlcolor=blue,citecolor=blue,linkcolor=blue]{hyperref}
\usepackage{graphicx}
\usepackage{multirow}
\usepackage{fancyhdr}
\def \solar{_\odot}
\def \pow10#1{\times 10^{#1}}

\tighten

\begin{document}

\submitted{Submitted to ApJ}

\title{Galaxy environments over cosmic time: 
  the non-evolving radial galaxy distributions around massive galaxies since 
  $\lowercase{z}=1.6$} 

\author{Tomer Tal\hyperlink{afflink}{$^{1}$}}
\author{Pieter G. van Dokkum\hyperlink{afflink}{$^2$}}
\author{Marijn Franx\hyperlink{afflink}{$^3$}}
\author{Joel Leja\hyperlink{afflink}{$^2$}}
\author{David A. Wake\hyperlink{afflink}{$^{4}$}}
\author{Katherine E. Whitaker\hyperlink{afflink}{$^{5}$}}

\affiliation{\hypertarget{afflink}{$^1$UCO/Lick Observatory, University of 
    California, Santa Cruz, CA 95064, USA;
    \href{mailto:tomer.tal@yale.edu}{tal@ucolick.org}}\\
  \hypertarget{afflink}{$^2$Yale University Astronomy Department, 
    P.O. Box 208101, New Haven, CT 06520-8101 USA}\\
  \hypertarget{afflink}{$^3$Leiden Observatory, Leiden University, 
    NL-2300 RA Leiden, The Netherlands}\\
  \hypertarget{afflink}{$^4$Department of Astronomy, University of 
    Wisconsin-Madison, Madison, WI 53706, USA}\\
  \hypertarget{afflink}{$^5$Astrophysics Science Division, Goddard 
    Space Flight Center, Greenbelt, MD 20771, USA}}

\begin{abstract}
  We present a statistical study of the environments of massive  
  galaxies in four redshift bins between $z=0.04$ and $z=1.6$, using data from 
  the Sloan Digital Sky Survey (SDSS) and the NEWFIRM Medium Band Survey (NMBS).
  We measure the projected radial distribution of galaxies in cylinders around 
  a constant number density selected sample of massive galaxies and utilize
  a statistical subtraction of contaminating sources.
  Our analysis shows that massive primary galaxies typically live in group 
  halos and are surrounded by 2 to 3 satellites with masses more than 
  one-tenth of the primary galaxy mass.
  The cumulative stellar mass in these satellites roughly equals the mass 
  of the primary galaxy itself.
  We further find that the radial number density profile of galaxies around 
  massive primaries has not evolved significantly in either slope or overall 
  normalization in the past 9.5 Gyr.
  A simplistic interpretation of this result can be taken as evidence for 
  a lack of mergers in the studied groups and as support for a static 
  evolution model of halos containing massive primaries.
%  We test this hypothesis by calculating the effect of mergers on the number 
%  density profile evolution and conclude that even a low merger rate would 
%  dramatically alter the derived galaxy profiles.
%  Nevertheless, observational studies and numerical calculations find that 
%  mergers play a major role in the evolution of massive quiescent galaxies. 
  Alternatively, there exists a tight balance between mergers and accretion 
  of new satellites such that the overall distribution of galaxies in and
  around the halo is preserved.
  The latter interpretation is supported by a comparison to a semi-analytic 
  model, which shows a similar constant average satellite distribution over 
  the same redshift range.\\

\end{abstract}


\section{\label{intro}Introduction}
 The properties of galaxies in the nearby universe are strongly
 affected by the environment in which they reside.  
 Morphology, mass, star formation rates and stellar colors of individual 
 galaxies have all been shown to be correlated with the local galaxy
 density (e.g., Oemler \citeyear{oemler_systematic_1974}; 
 Dressler \citeyear{dressler_galaxy_1980};
 Kauffmann et al. \citeyear{kauffmann_environmental_2004};
 Hogg et al. \citeyear{hogg_dependence_2004}; 
 Blanton et al. \citeyear{blanton_relationship_2005}; 
 Thomas et al. \citeyear{thomas_epochs_2005}; 
 Clemens et al. \citeyear{clemens_star_2006};
 van den Bosch et al. \citeyear{van_den_bosch_importance_2008};
 Skibba \& Sheth \citeyear{skibba_halo_2009}; 
 Peng et al. \citeyear{peng_mass_2010};
 Zehavi et al. \citeyear{zehavi_galaxy_2011};
 Quadri et al. \citeyear{quadri_tracing_2012}).
 Therefore, significant effort has been devoted to studies of galaxy 
 environments at low and high redshift, utilizing a number of approaches 
% for estimating local galaxy clustering 
 (e.g., Zehavi et al. \citeyear{zehavi_galaxy_2002}; 
 Madore et al. \citeyear{madore_companions_2004}; 
 Mandelbaum et al. \citeyear{mandelbaum_galaxy_2006};
 Holden et al. \citeyear{holden_mass_2007}
 Gavazzi et al. \citeyear{gavazzi_sloan_2007}; 
 Bolton et al. \citeyear{bolton_sloan_2008}; 
 Wake et al. \citeyear{wake_2df-sdss_2008}, \citeyear{wake_galaxy_2011};
 Patel et al. \citeyear{patel_dependence_2009},
 \citeyear{patel_star-formation-rate-density_2011};
 Cacciato et al. \citeyear{cacciato_cosmological_2012};
 Knobel et al. \citeyear{knobel_zcosmos_2012};
 Diener et al. \citeyear{diener_proto-groups_2012}).
 Nevertheless, the density and distribution of galaxies in groups, where nearly 
 all massive galaxies are expected to reside, are still poorly known at $z>1$. 
 At this redshift, it is difficult to uniquely match galaxies to halos except 
 in the most overdense regions.

 Recently, a number of studies analyzed the small-scale clustering
 of satellite galaxies around massive primaries (the most massive 
 galaxies in their halos) using statistical tools.
 By utilizing large data sets and statistical background subtraction, authors 
 were able to extract satellite galaxy distributions from 
 observational and numerical catalogs 
 (e.g., Masjedi et al. \citeyear{masjedi_very_2006}, 
 \citeyear{masjedi_growth_2008};
 Watson et al. \citeyear{watson_extreme_2012};
 Tal et al. \citeyear{tal_observations_2012};
 Quilis \& Trujillo \citeyear{quilis_satellites_2012}; 
 Budzynski et al. \citeyear{budzynski_radial_2012}; 
 Neirenberg et al. \citeyear{nierenberg_luminous_2012};
 Jiang et al. \citeyear{jiang_distribution_2012};
 Wang \& White \citeyear{wang_satellite_2012};
 M\'{a}rmol-Queralt\'{o} et al. \citeyear{marmol-queralto_characterizing_2013};
 Guo et al. \citeyear{guo_spatial_2013}).
 
 Here we follow a similar approach and derive galaxy number-density
 profiles out to $z=1.6$ to study the properties and evolution of galaxy 
 distributions around massive primaries.
 An important advantage of this approach is that it does not rely on accurate 
 redshift measurements of the studied galaxies. 
% and it benefits from the relative abundance of galaxy group halos.
 
 Throughout the paper we adopt the following cosmological parameters:
 $\Omega_m=0.3$, $\Omega_{\Lambda}=0.7$ and $H_0=70$ km s$^{-1}$
 Mpc$^{-1}$.
 
\section{Sample selection}
 \label{sec:data}
 We utilized data from two public surveys for this study; from SDSS 
 (York et al. \citeyear{york_sloan_2000}) and from NMBS 
 (van Dokkum et al. \citeyear{van_dokkum_newfirm_2009}; 
 Whitaker et al. \citeyear{whitaker_newfirm_2011}).
 Galaxies in three redshift bins in the range $0.4<z<1.6$ were selected 
 from NMBS, a deep imaging program with the NOAO Extremely Wide-Field 
 Infrared Imager.
 The total imaging area of NMBS spans roughly 0.39 deg$^2$ over two
 fields and the catalogs include tens of thousands of galaxies for
 which excellent photometric redshifts were calculated 
 ($\sigma_z/(1+z)\sim0.02$; Brammer et al. \citeyear{brammer_dead_2009}).
 Low redshift galaxies were selected at $0.04<z<0.07$ from the MPA-JHU 
 emission line analysis 
 catalogs\footnote{http://www.mpa-garching.mpg.de/SDSS/DR7/} which 
 are based on the seventh data release of SDSS 
 (Abazajian et al. \citeyear{abazajian_seventh_2009}).
% Redshifts of all candidate SDSS galaxies were determined
% spectroscopically as part of the survey over an area of more than
% 7600 deg$^2$.

 \subsection{Cumulative number density matching}
  Studies of galaxy evolution often match their samples such that galaxies 
  at all redshifts have similar observed properties 
  (e.g., constant stellar mass, luminosity).
  However, the masses and luminosities of galaxies are expected to evolve and
  change with time.
  Alternatively, samples can be matched based on their number density, 
  regardless of galaxy properties.
  Since cumulative number density is not expected to change dramatically even
  with a non-negligible number of mergers 
  (e.g., Gao et al. \citeyear{gao_early_2004}; 
  van Dokkum et al. \citeyear{van_dokkum_growth_2010}; 
  Papovich et al. \citeyear{papovich_rising_2011};
  Patel et al. \citeyear{patel_hst/wfc3_2012};
  Leja et al. submitted), this approach essentially follows the evolution 
  of observed properties of a given galaxy population.
  We utilized the latter technique to match massive galaxy populations
  in the two data sets in four redshift bins.

 \begin{figure}
   \includegraphics[width=0.48\textwidth]{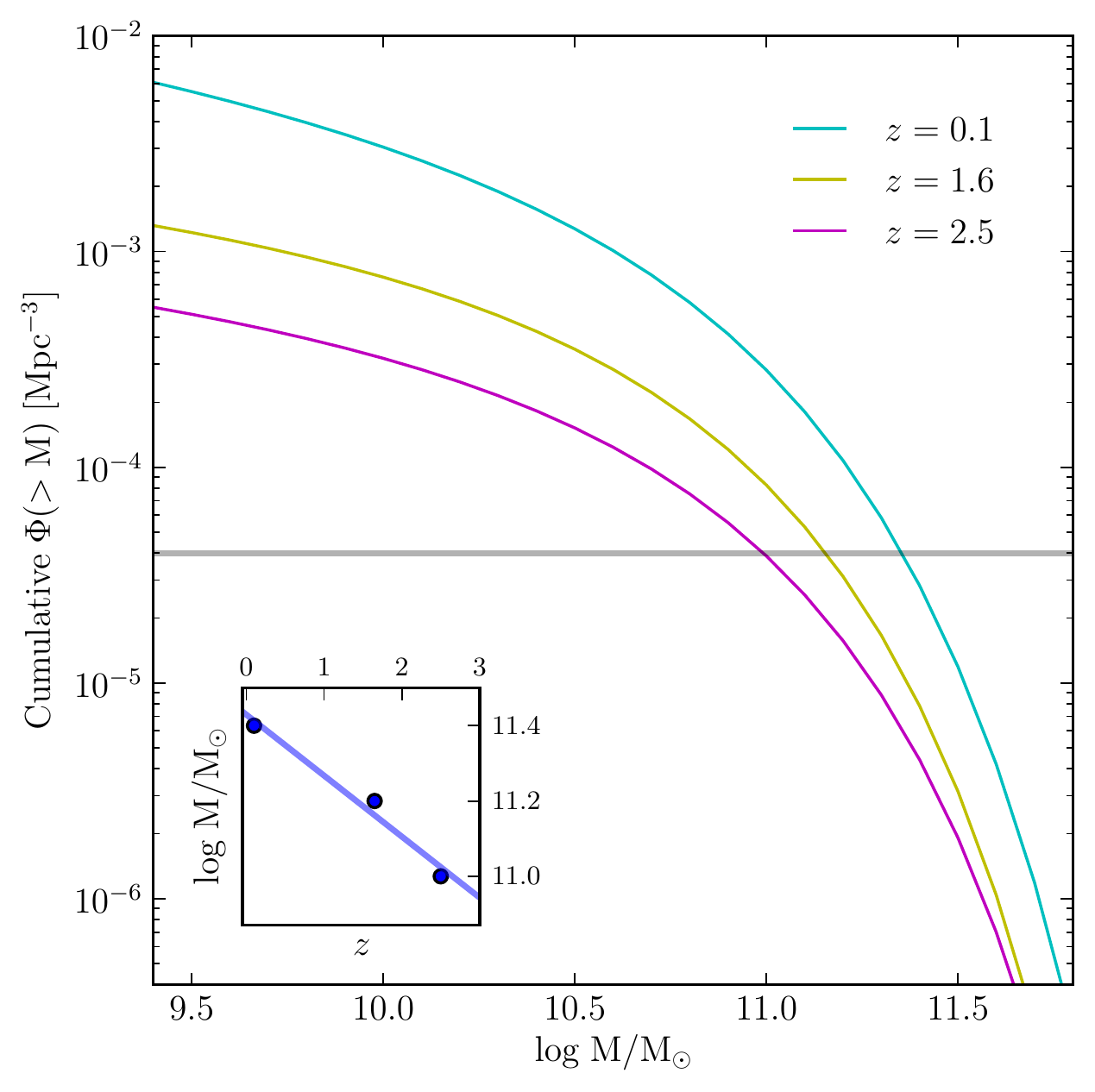}
   \caption{Sample matching of the four redshift bins.
     The teal, yellow and magenta lines represent the cumulative
     mass functions of Marchesini et
     al. (\citeyear{marchesini_evolution_2009}).
     We calculated galaxy masses at a constant cumulative number
     density of 4$\pow10{-5}$ Mpc$^{-3}$ (gray line) and derived a
     mass-redshift relation (inset figure).
     We then selected galaxies in $\log M$ bins according to the derived 
     relation.
   }
   \label{fig:m09_sel}
 \end{figure}

  We started by adopting the model fits to the mass functions from
  Marchesini et al. (\citeyear{marchesini_evolution_2009}) to
  calculate the cumulative number density of three redshift samples as
  a function of stellar mass (Figure \ref{fig:m09_sel}).
  We then fit a line to the mass-redshift relation of galaxies at a
  constant number density of 4$\pow10{-5}$ Mpc$^{-3}$ and found 
  the following mass evolution with redshift:
  \begin{equation}
    \log(M_\star/M\solar)=-0.16z+11.43
    \label{eq1}
  \end{equation}
  The best-fit relation implies doubling of stellar mass in the last 9.5 
  Gyr for these galaxies and it is consistent with other mass 
  evolution studies (e.g., van Dokkum et al. \citeyear{van_dokkum_growth_2010}).
  We selected primary galaxy candidates in $\log{M_\star}$ bins 
  of width 0.3 dex with an evolving median mass according to Equation 
  \ref{eq1}.
  Galaxies were considered to be ``primary'' if no other, more massive, 
  galaxies were found within a projected radius of 500 kpc.
  Otherwise, they were counted as ``satellites'' of their more massive
  neighbor.
  The redshift limits for this study were determined such that the samples
  from both surveys were complete down to one-tenth of the stellar mass of
  all selected primaries.
  Redshift bin sizes were chosen to have roughly equal volumes in the
  three NMBS bins and to maximize the number of galaxies from SDSS, whilst 
  remaining complete in stellar mass.

  A summary of the selected samples is given in Table \ref{tab:bins}.

  \begin{table}[t]
    \caption{Galaxy properties in the four selected redshift bins.}
    \centering
    \begin{tabular}{l c c c c}
      \hline\hline
      Data set & $z$ & V [Mpc$^3$] & log(M$_{med}$/M$\solar$) &
      N$_\mathrm{P}$ \\
      \hline {\vspace{-5px}}\\
      SDSS & 0.04$<$$z$$<$0.07 & 1.6$\pow10{7}$ & 11.42 & 360 \\
      \multirow{3}{*}{NMBS} & 0.4$<$$z$$<$0.9 & 9.8$\pow10{5}$ & 11.33
      & 44 \\
      & 0.9$<$$z$$<$1.3 & 1.4$\pow10{6}$ & 11.25 & 89 \\
      {\vspace{3px}}& 1.3$<$$z$$<$1.6 & 1.2$\pow10{6}$ & 11.20 & 97\\
      \hline\hline
      \label{tab:bins}      
    \end{tabular}
  \end{table}
  
\section{Radial number density profiles}
\label{sec:derive}
 Various techniques are commonly used to quantify galaxy environments at 
 low and high redshift.
 Here we perform an analysis of the distribution of objects around the 
 selected galaxies and subtract the contribution of background 
 and foreground sources in a statistical manner.
 One of the advantages of this method is that it does not rely on 
 a priori assumptions regarding the total mass or size of the host
 dark matter halo.

 \begin{figure*}
   \includegraphics[width=1.0\textwidth]{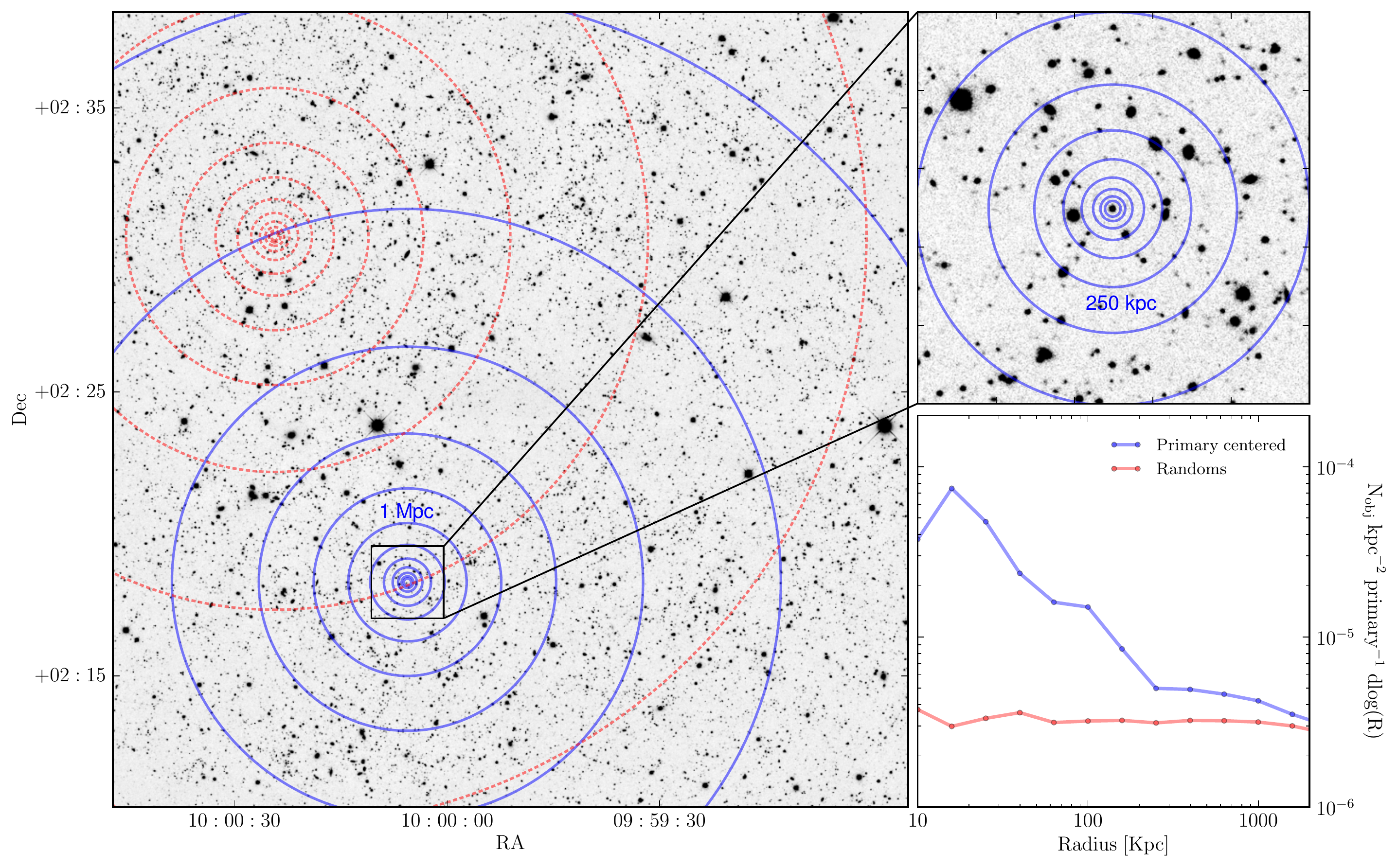}
   \caption{Demonstration of the radial number density profile extraction
     technique.
     We divided catalog sources into $\log(r)$ bins around the selected
     primaries (blue lines) and around randomly selected positions in
     the field (red lines).
     The bottom-right panel shows an example of the stacked radial
     profiles of all galaxies at $1.3<z<1.6$.
     The projected radial profiles of galaxies around the
     selected primaries were revealed by subtracting the average random 
     profile from the galaxy centered profile.
   }
   \label{fig:examp}
 \end{figure*}
 
 \subsection{Profile derivation}
 \label{sec:pderive}
  We started by selecting all galaxies in the NMBS catalog that were
  within $\Delta z\le0.2$ of the measured redshift of the massive primaries and 
  within $\Delta z\le0.005$ in the SDSS MPA-JHU catalog.
  From this selection we excluded all galaxies whose mass was less than
  one-tenth of their corresponding primary mass.
  We then derived the radial number density galaxy profiles by binning 
  the selected galaxies in log-radial bins 
  (blue lines in Figure \ref{fig:examp}).
  
  The resulting radial number density profiles include galaxies that are 
  physically associated with each primary as well as sources in the 
  background and foreground.
  In order to account for this contamination we repeated the derivation 
  in randomly selected positions within the entire survey area and calculated 
  the average radial profile of contaminating sources (red lines in Figure 
  \ref{fig:examp}).
  Finally, we subtracted the average random profile from the average 
  primary profile at each redshift to derive the cylindrical distribution of 
  physically associated galaxies around massive primaries.

  The radial distribution of galaxies around massive primaries out to 
  $z=1.6$ is shown in Figure \ref{fig:profiles}, where the density of 
  physically associated galaxies is plotted as a function of projected 
  physical distance from the primary.
  The blue line represents the measurement at $0.04<z<0.07$ from 
  SDSS while the yellow, green and red lines show the three higher 
  redshift bins from NMBS.
  We note that the lowest redshift profile (blue line) is not plotted at 
  $r<55''$ (roughly 74 kpc at $z=0.07$), where the spectroscopic catalog
  of SDSS is incomplete due to fiber collisions.\\

 \subsection{Error estimates}
  The range of measured number density values in the random profiles 
  reflects the statistical variation of the overall galaxy distribution 
  throughout the survey fields.
  In addition to averaging all random profiles in each redshift bin, we 
  also calculated the standard deviation of the distribution of individual 
  profiles as a function of radius.
  This measurement depicts the statistical uncertainty in our calculations.

  In addition to statistical errors, a number of systematic uncertainties 
  may potentially affect our results in a non-trivial way.
  For example, uncertainties in mass and redshift measurements may move 
  galaxies in and out of a given redshift sample and therefore change the 
  profile normalization and slope.
  In the innermost radii the profiles may be underestimated
  due to source blending close to the bright primaries.
  While redshift errors in NMBS are well quantified and are expected to have 
  a minor effect on the number density profiles, uncertainties in stellar mass 
  measurements are complex and depend on a large number of fitting parameters 
  (e.g., Marchesini et al. \citeyear{marchesini_evolution_2009}; 
  Muzzin et al. \citeyear{muzzin_near-infrared_2009}; 
  Conroy et al. \citeyear{conroy_propagation_2009}).

  We quantified the effect of redshift errors by scattering all 
  measured redshifts in the NMBS catalog by a normal distribution with width 
  $\sigma_z/(1+z)=0.02$. 
  This value represents the scatter in the relation between photometric and 
  spectroscopic redshift measurements in NMBS as was found by Brammer et al. 
  (\citeyear{brammer_dead_2009}).
  We extracted number density profiles using the scattered catalog, 
  following the procedure described in Section \ref{sec:pderive}.
  The overall shape and normalization of the resulting profiles are 
  consistent with the non-scattered profiles and the resulting errors are 
  within $\pm20\%$ of the above derived statistical errors at all radii.
  The error bars in Figure \ref{fig:profiles} represent statistical 
  uncertainties as well as redshift errors in this study.

  Measured values and error estimates that are presented in 
  Figure \ref{fig:profiles} are given in Table \ref{tab:profiles}.

 \subsection{Profile dependence on primary mass}
\begin{figure}
    \includegraphics[width=0.48\textwidth]{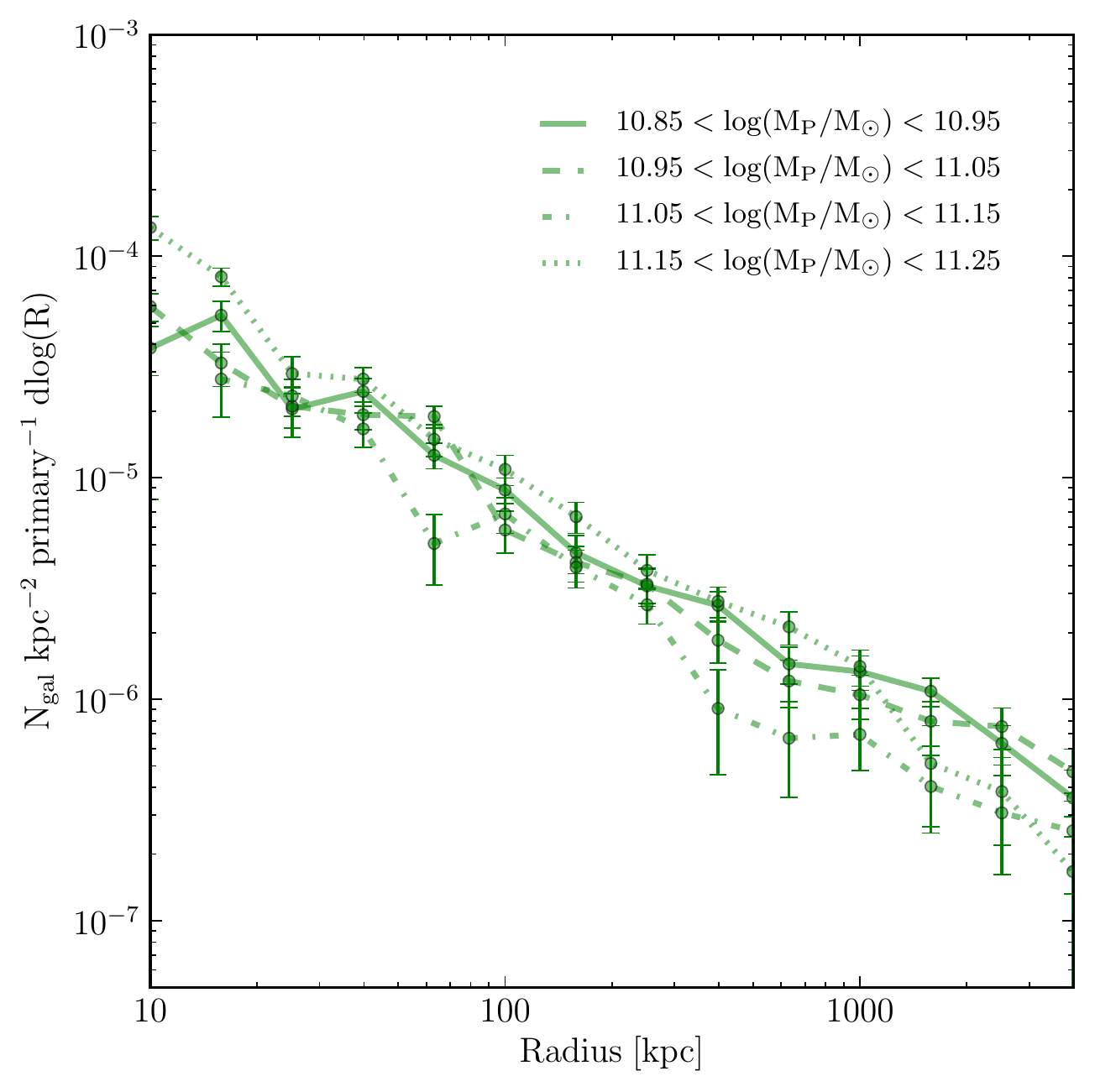}
    \caption{Projected radial number density profiles in the redshift range 
      $0.9<z<1.3$ in four mass bins.
      We found no evidence for a strong dependence on the primary galaxy 
      mass within the probed mass range.
    }
    \label{fig:constz}
  \end{figure}

  In order to test whether our results are sensitive to the mass of the 
  primaries we divided the $0.9<z<1.3$ sample into four bins 
  according to the total stellar mass of the primary galaxies.
  We selected primaries with masses in the range 
  $10.85<\log{(M_P/M\solar)}<11.25$ and calculated the radial number density 
  profile of each mass bin following the method described in Section 
  \ref{sec:derive}.
  We estimated the statistical uncertainties for each mass bin using the
  variation in individual profile measurements.
  Figure \ref{fig:constz} shows that the radial number density profiles of all 
  four mass bins are consistent with one another and no strong mass
  dependence is apparent in this sample.
  However, we note that the relatively small mass range of selected 
  primaries may be insufficient to detect even a moderate mass dependence.\\
\\
 \subsection{Environment of massive primaries}
\begin{figure}
    \includegraphics[width=0.48\textwidth]{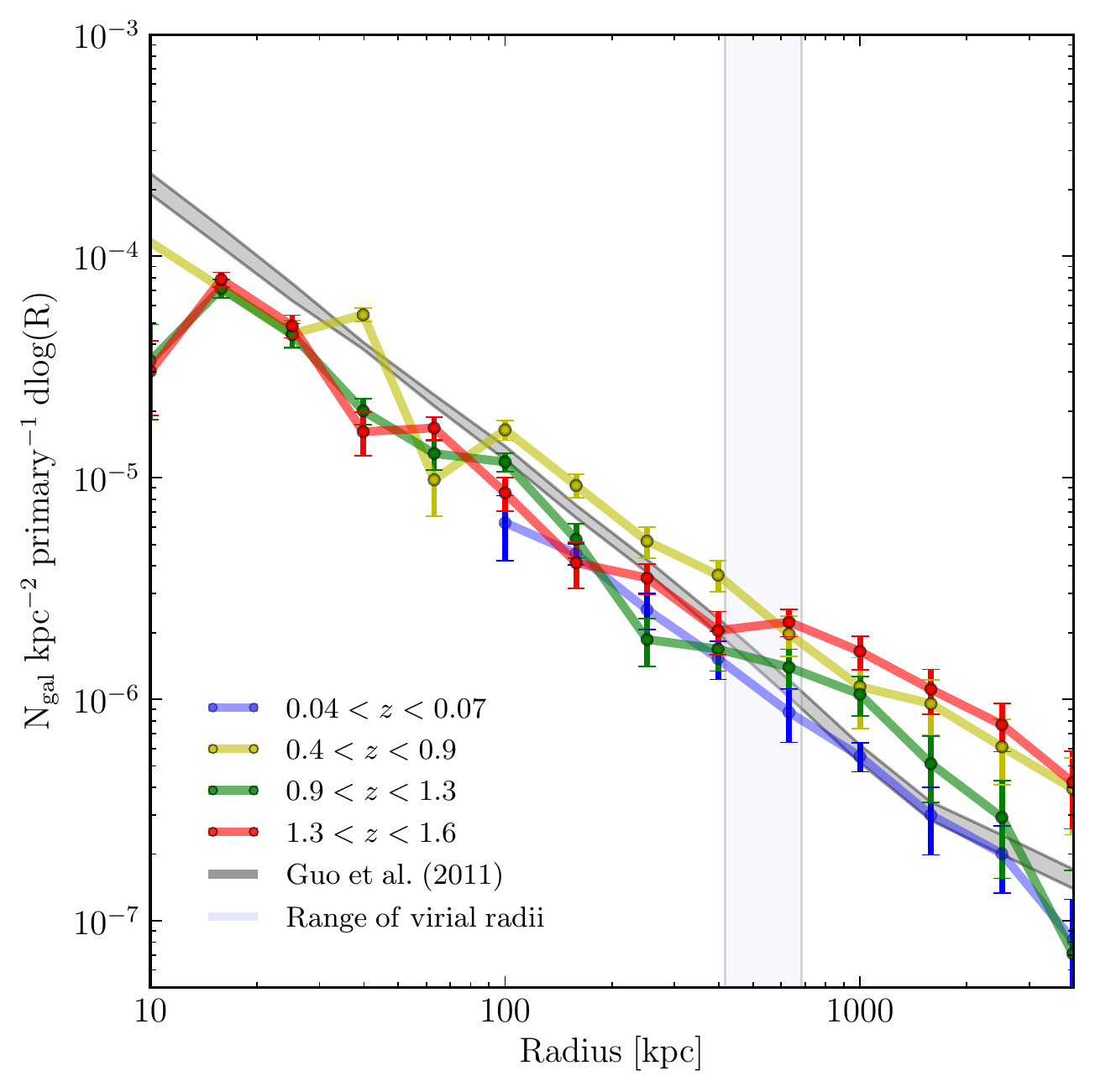}
    \caption{Projected radial profile of galaxies around the selected 
      primaries in four redshift bins (colored lines).
      The profiles show a remarkable lack of evolution in the radial
      galaxy distribution since $z=1.6$.
      The shaded gray profile represents a similar derivation
      using the semi-analytic model of Guo et al. (\citeyear{guo_dwarf_2011}), 
      over the same redshift range.
      The vertical light gray area shows the range of virial radius values 
      of the dark matter halos containing the modeled galaxies.
      The observed profiles are consistent with the modeled ones over
      most of the radial range.
    }
    \label{fig:profiles}
  \end{figure}

\begin{figure}
    \includegraphics[width=0.48\textwidth]{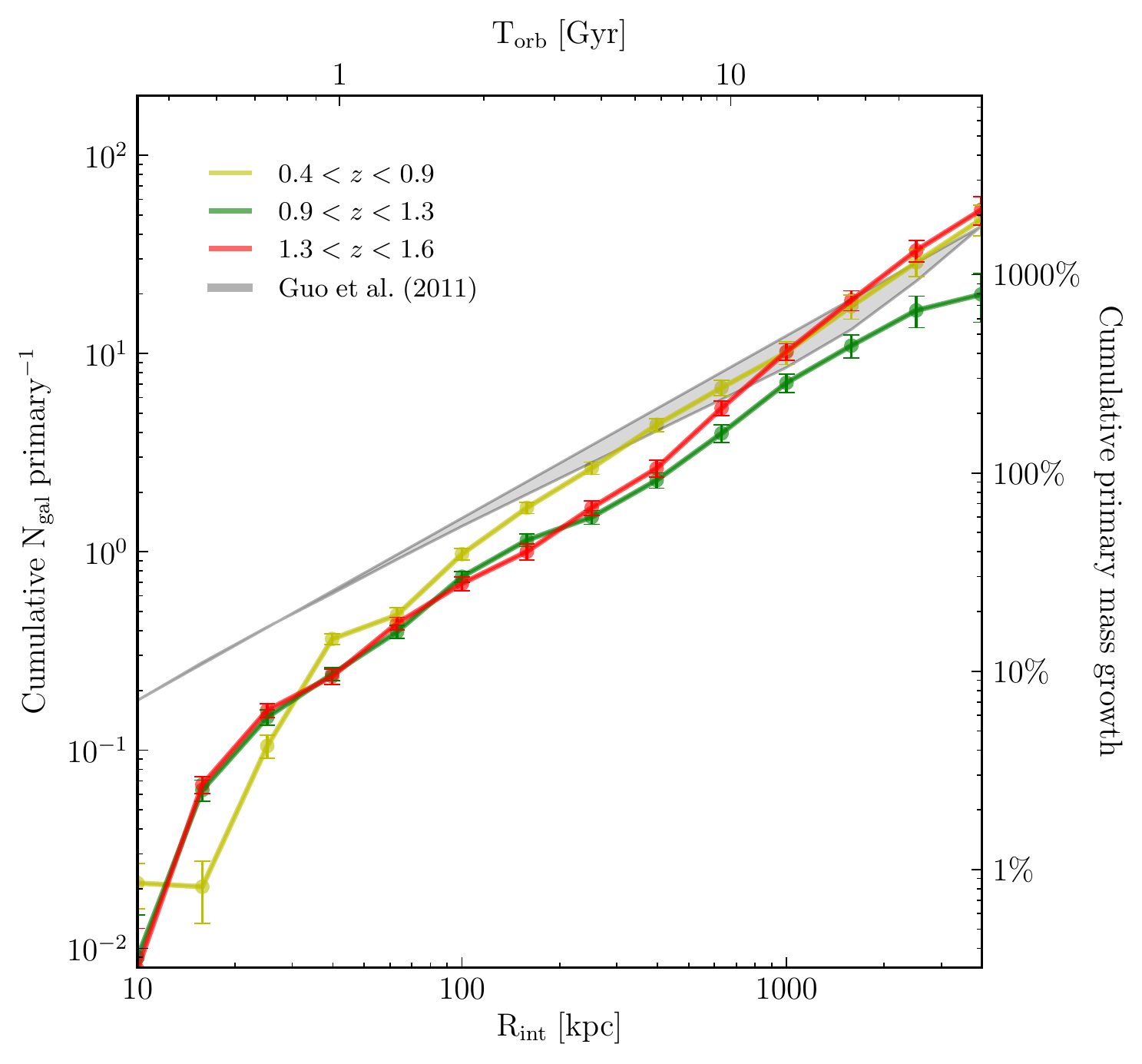}
    \caption{Cumulative number of galaxies as a function of
      integration radius from NMBS (solid yellow, green and red
      lines) and from G11 (gray shaded region).
      The upper x-axis shows an estimate of the orbital time scale
      under an assumed NFW halo with total mass $2\pow10{13}\
      M\solar$.
      The right-hand y-axis displays the radially integrated galaxy
      mass, assuming that all galaxies have the median mass of 40\%
      of their respective primary mass.
      Primary galaxies can double in mass by merging with 2.5  
      galaxies that are at $r\lesssim300$ kpc.
    }
    \label{fig:inttot}
  \end{figure}

  In order to calculate the mass density in the vicinity of our sample 
  primaries we converted the number density profiles from Figure 
  \ref{fig:profiles} into a cumulative number of satellites as a function 
  of radius.
  Figure \ref{fig:inttot} shows the integrated curves of the NMBS  
  profiles as a function of integration radius.
  The top x-axis of the Figure shows the orbital timescale 
  of a test particle in a spherical halo with total mass $2\pow10{13}$ 
  M$\solar$ which is distributed as a Navarro, Frenk \& White 
  (\citeyear{navarro_universal_1997}; NFW) profile.
  The y-axis on the right-hand side of Figure \ref{fig:inttot} 
  represents an estimate of the total enclosed mass as a function of radius,
  assuming that the mass of each plotted galaxy equals 40\% of its primary 
  mass.
  This value represents the median mass of all selected galaxies as calculated 
  from the global mass function at $z=1.6$.

  Average properties of massive galaxy environments out to $z=1.6$ can be
  directly measured from these integrated profiles.
  For example, these profiles can be directly compared with estimates of 
  galaxy pair fractions
  (e.g., Le F\`{e}vre et al. \citeyear{le_fevre_hubble_2000}; 
  Patton et al. \citeyear{patton_dynamically_2002}; 
  Lin et al. \citeyear{lin_deep2_2004};
  Bell et al. \citeyear{bell_merger_2006}; 
  Kartaltepe et al. \citeyear{kartaltepe_evolution_2007};
  Bluck et al. \citeyear{bluck_surprisingly_2009}; 
  Bundy et al. \citeyear{bundy_greater_2009};
  L\'{o}pez-Sanjuan et al. \citeyear{lopez-sanjuan_dominant_2012}).
  Recently, Williams et al. (\citeyear{williams_diminishing_2011}) and 
  Newman et al. (\citeyear{newman_can_2012}) found that roughly 20\% of 
  all massive galaxies have a luminous companion with a mass ratio of up 
  to 10:1 and within a projected distance of 30 kpc.
  The same result can be directly extracted from Figure \ref{fig:inttot} 
  by reading off an average value of 0.2 galaxies within 30 kpc of the 
  primaries.

  In addition to quantifying pair fractions in discrete separation values, 
  the derived cumulative profiles describe the enclosed mass in a range of 
  radii, essentially quantifying the average stellar mass content of studied 
  halos.
  The total number of galaxies within a mass range of 1:10 and within roughly 
  400 kpc of the primary is on average 2 to 3 in all redshift bins.
  This value suggests that massive primary galaxies have typically resided 
  in group halos in the last 9.5 Gyr and it is comparable to the number of 
  luminous satellites in massive groups from other studies (e.g., 
  Tal et al. \citeyear{tal_observations_2012};
  Quilis \& Trujillo \citeyear{quilis_satellites_2012}),  
  Moreover, the total stellar mass in satellites within the halo virial radius
  can be inferred from the cumulative profiles and it roughly equals the mass 
  of the primary galaxy itself.

 \subsection{Profile redshift evolution} 
  The average radial profiles of the four redshift bins show a remarkable 
  lack of evolution in shape and overall normalization.
  Within the measurement scatter, the profiles are consistent with 
  one another on nearly all scales from $z=1.6$ to $z\sim0$.
  The lack of evolution in the number density profiles can be naively taken 
  as evidence for a static model of massive halo evolution.
  According to the stable clustering model (Davis \& Peebles 
  \citeyear{davis_integration_1977}), cluster halos decouple from the 
  large-scale cosmological flow shortly after formation.
  By that point, the halos are virialized and they show no significant 
  evidence for radial infall.
  As a consequence, any mergers that involve primary galaxies can only be  
  with other galaxies from the same halo.

\begin{figure}
   \includegraphics[width=0.421\textwidth]{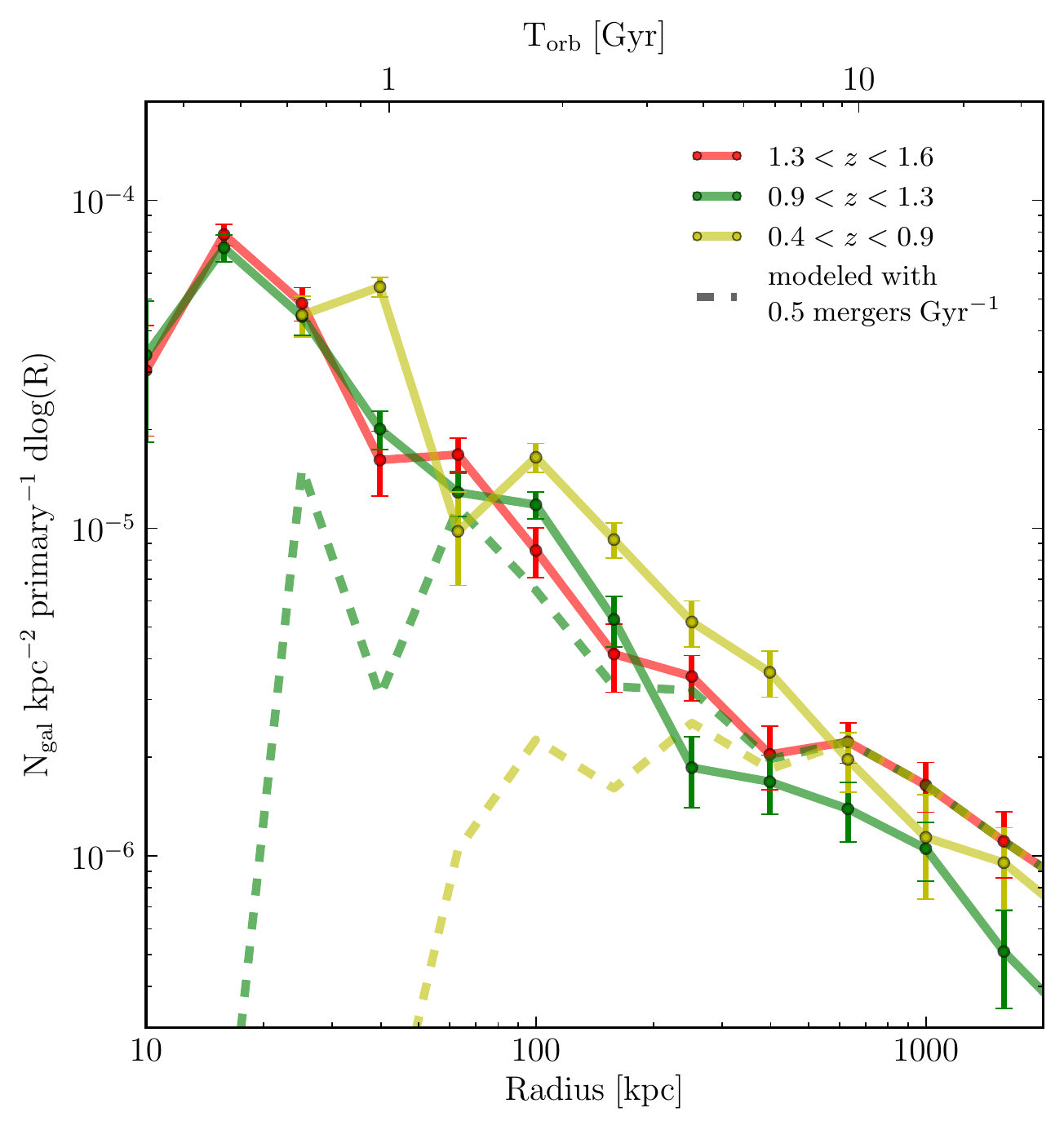}
   \caption{Estimates of the effect of in-halo mergers on the radial
     number density profile evolution.
     We modeled mergers by excluding galaxies from the derived profiles 
     of each primary galaxy in the range $10<r/\mathrm{kpc}<400$.
     The dashed lines represent the expected evolution of the 
     $1.3<z<1.6$ profile into subsequent redshift bins at a merger rate 
     of 0.5 Gyr$^{-1}$ with no external halo accretion.
   }
   \label{fig:mergerz}
 \end{figure}

  In order to test the effect of in-halo mergers on the evolution of the 
  radial number density profile we simulated mergers starting with the 
  highest redshift bin ($1.3<z<1.6$) and compared the results to the measured 
  profiles at lower redshifts.
  Mergers were simulated by excluding galaxies from the radial galaxy
  distributions in the range $10<r/\mathrm{kpc}<400$, essentially allowing
  satellites to merge with the primary from anywhere in the halo.
  For an estimate of the average merger rate we utilized the results from 
  van Dokkum et al. (\citeyear{van_dokkum_growth_2010}), who found that 
  massive primary galaxies have roughly doubled in stellar mass since $z=2$.
  Assuming that most of the mass growth takes place via minor mergers with 
  mass ratios less than 10:1 (Tal et al. \citeyear{tal_mass_2012}), the 
  implied merger rate is roughly 0.5 Gyr$^{-1}$.
  Figure \ref{fig:mergerz} shows the modeled profiles at the two 
  lowest NMBS redshifts (dashed lines), as well as the observed profiles
  (solid lines).

  It is clear from Figure \ref{fig:mergerz} that in-halo mergers 
  would have a pronounced effect on the evolution of the radial number
  density profile.
  If each primary galaxy underwent mergers at the estimated rate and no 
  new galaxies were added to the halos on similar scales, the observed 
  profile at $1.3<z<1.6$ (solid red line) would evolve into the modeled 
  profiles at $0.9<z<1.3$ (dashed green line) and at $0.4<z<0.9$ 
  (dashed yellow line).
  Instead, the observed profile at these redshifts (solid green and yellow 
  lines) are consistent with no evolution at all.
  
  Taken at face value, this analysis supports a static model, suggesting 
  that a significant rate of galaxy mergers within the halo is unlikely 
  on time scales of up to roughly 10 Gyr.
  However, studies of the stellar mass function of massive galaxies show that
  such galaxies grow significantly in mass over the same redshift range, 
  mainly through mergers
  (e.g., Dickinson et al. \citeyear{dickinson_evolution_2003}; 
  Bundy et al. \citeyear{bundy_mass_2006}; 
  Drory et al. \citeyear{drory_stellar_2005}; 
  Pozzetti et al. \citeyear{pozzetti_vimos_2007}; 
  P\'{e}rez-Gonzalez et al. \citeyear{perez-gonzalez_stellar_2008}; 
  Marchesini et al. \citeyear{marchesini_evolution_2009}).
  It is therefore evident that truly non-evolving models insufficiently
  recover the observed evolution of massive primary galaxies and their 
  environments.

\section{Comparison to a semi-analytic model}
 \subsection{Density profiles and mass growth}
  In order to study the evolution of massive galaxy groups in the
  last 9 Gyr we examined numerical predictions using   
  the semi-analytic model of Guo et al. (\citeyear{guo_dwarf_2011}; G11).
  G11 applied a semi-analytic model to the merger trees of the Millennium 
  Simulation (Springel et al. \citeyear{springel_simulations_2005}) and 
  improved the treatment of gas dynamics and tidal disruptions compared to 
  previous studies.

  For consistency, we first test whether our cumulative number density 
  matching technique indeed identifies progenitors and descendants in G11.
  We selected modeled galaxies at $z=1.63$ using our cumulative number 
  density selection criteria and followed their stellar mass evolution at all 
  simulated time points down to $z=0$.
  The median growth factor of massive primary galaxies in G11 is roughly 2.4,
  consistent with the growth that was measured observationally for similarly 
  massive galaxies by van Dokkum et al. (\citeyear{van_dokkum_growth_2010}).
  We also followed the simulated mass evolution of the group dark matter 
  halos and compared it to the average ratio between the halo mass of 
  galaxies selected at $z=0$ and at $z=1.63$.
  We found that both simulated and predicted mass growth factors are within
  4\% of one another with an average halo growth by a factor of roughly 3.7.
%  We used the G11 models and analyze descendants of the halos selected by 
%  our cumulative density criterion at z=1.6. 
%  Their descendant masses at z=1 and 0,.6 are XXX and YYY, very similar to 
%  the typical halos at ncum= XXX which have halo masses of ZZZ and UUUU. 
  In short, by matching massive galaxies by their cumulative number density 
  we were able to correctly trace progenitor and descendant halos.

  From the galaxy catalogs of G11 we selected all primary galaxies at 
  redshifts $z=$0.06, 0.62, 1.17 and 1.63, following the same selection
  criteria as those used for the observed data.
  We then collapsed the simulation box along one of the physical dimensions
  and extracted the projected radial distributions of physically associated 
  galaxies around the selected primaries following the method described in 
  Section \ref{sec:data}.
  This approach allows for a direct comparison with the observational results, 
  as the uncertainties associated with the profile extraction procedure are 
  preserved.
  The gray profile in Figure \ref{fig:profiles} represents this measurement
  in the full redshift range, $0.06<z<1.63$.
  The vertical light-gray region in the Figure shows the range of virial 
  radius values of the dark-matter halos in which all selected primaries
  reside, as determined by G11.
  Finally, the gray curve in Figure \ref{fig:inttot} shows the integrated 
  number density profiles over the same redshift range.
  It is evident from Figure \ref{fig:profiles} that the average modeled 
  and observed profiles agree well with each other on all but the 
  smallest scales.
  We note that the innermost observational data point (at r$<$10 kpc) is 
  likely underestimated since galaxies inside this radius are within the 
  full-width half maximum of the average point spread function at $z>1$.

%  As a consistency check we measured the mass growth of massive primary 
%  galaxies selected at $z=1.63$ from G11 by following their stellar mass 
%  at all simulated time points down to $z=0$.
%  The median growth factor of massive primary galaxies in G11 is roughly 2.4,
%  consistent with the growth that was measured observationally for similarly 
%  massive galaxies by van Dokkum et al. (\citeyear{van_dokkum_growth_2010}).
%  We also followed the simulated mass evolution of the group dark matter 
%  halos and compared it to the average ratio between the halo mass of 
%  galaxies selected at $z=0$ and at $z=1.63$.
%  We found that both simulated and expected mass growth factors are within
%  4\% of one another with an average halo growth by a factor of roughly 3.7.

  \begin{figure}
   \includegraphics[width=0.48\textwidth]{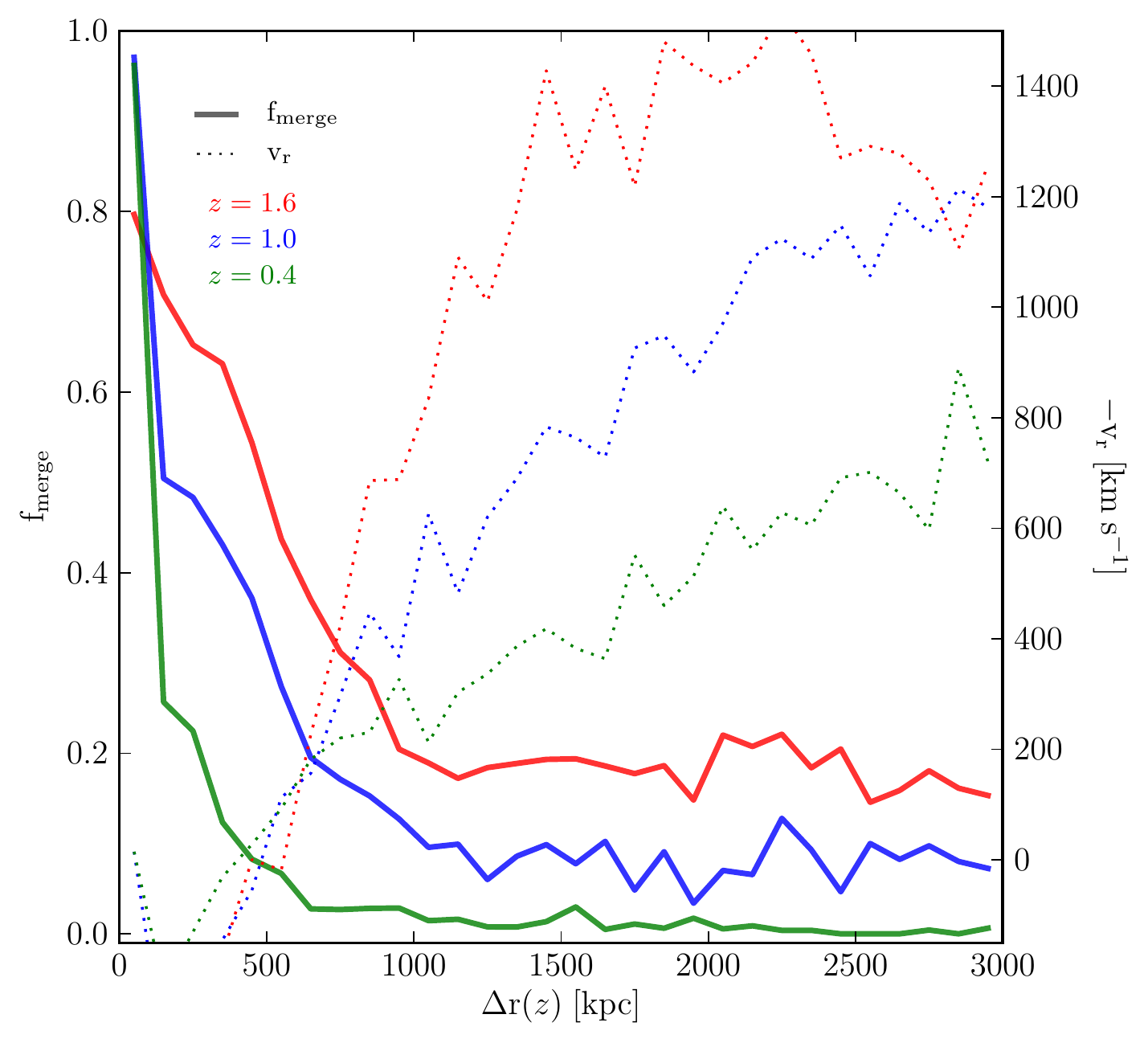}
   \caption{Fate of galaxies around massive primaries in G11.
     Solid lines represent the fraction of modeled galaxies that merge with
     their respective primary by $z\sim0$ as a function of initial 
     separation.
     Dotted lines show the derived radial component of the velocity vector
     at the plotted redshift.
     Three initial redshift measurements, $z=1.6$, 1.0 and 0.4 are plotted
     in red, blue and green, respectively.
     Galaxies outside of 1000 kpc initially (roughly two virial radii) merge 
     with the primaries at a low rate of less than 20\% and move at a high
     radial velocity.
     Galaxies inside the halo virial radius merge with the primary at a 
     significantly higher rate with an average infall velocity that is close
     to zero.
   }
   \label{fig:fracvel}
 \end{figure}

 \subsection{Fate of modeled satellite galaxies}
  The good agreement between our result and the theoretical prediction 
  prompted us to further investigate the fate of galaxies around massive 
  primaries by following their merging histories and orbits in the models.
  To do so, we selected 1000 primary galaxies at $z=1.63$ from G11 at random, 
  following the selection criteria from Section \ref{sec:data}.
  We then traced the positions of galaxies in 3 Mpc apertures around the 
  primaries, that were more massive than one-tenth of the primary mass.
  We repeated the selection at all output redshifts in the range $0.06<z<1.63$ 
  and followed the trajectories of individual galaxies through time.
  In addition, we determined whether a satellite merges with its primary by 
  $z=0.06$ using the merger trees of the Millennium Simulation.

  The solid lines in Figure \ref{fig:fracvel} show the fraction of satellite 
  galaxies that eventually merge with their primary galaxy as a function of 
  physical separation at $z=1.6$, 1.0 and 0.4.
  Galaxies that are found inside the virial radius at any given redshift
  are significantly more likely to merge with their primary ($f_{merge}>60\%$ 
  at $z=1.6$) than galaxies that start off far from the halo ($f_{merge}<20\%$).
  However, as is shown in Figure \ref{fig:inttot}, only a small fraction 
  of all galaxies within 3 Mpc are found inside of the group virial radius 
  ($\sim$10\%).
  The y-axis on the right-hand side of Figure \ref{fig:fracvel} and 
  the dashed lines show the median radial velocity of satellite galaxies with 
  respect to their primaries, as a function of physical separation.
  At each redshift we estimated the average streaming velocities by measuring 
  the mean distance between satellites and primaries at two consecutive output 
  redshifts and dividing the difference between them by the total modeled time 
  span.  
  This measurement gives a rough estimates of the average radial component of 
  the velocity vector in the rest frame of the primary galaxy.  
  Figure \ref{fig:fracvel} shows that galaxies outside of roughly two virial 
  radii fall into the halo with high velocities.
%  At each redshift we calculated relative radial velocities by measuring the 
%  distance between satellites and primaries at two consecutive output 
%  redshifts and dividing the difference between them by the total modeled 
%  time span.
%  This measurement effectively estimates the average radial component of the 
%  velocity vector in the rest frame of the primary galaxy.
%  Figure \ref{fig:fracvel} shows that galaxies outside of roughly two virial 
%  radii fall into the halo with high velocities.

  The fate of satellite galaxies in the semi-analytic model of G11 seems to 
  be bimodal. 
  The vast majority of galaxies fall toward the most massive halo galaxy with 
  a high average radial velocity (in agreement with e.g., 
  Tormen \citeyear{tormen_rise_1997}; 
  Benson et al. \citeyear{benson_orbital_2005}; 
  Kochfar \& Burkert \citeyear{khochfar_orbital_2006}; 
  Wetzel \citeyear{wetzel_orbits_2011}) and do not merge with the primary 
  by $z\sim0$. 
  A small fraction of galaxies have a significantly lower mean radial 
  velocity and they eventually merge with the primary.  
  The probability that a galaxy will merge with its massive primary is 
  inversely proportional to the physical distance between them.  
  At any simulated redshift, most galaxies that are found outside of the 
  halo virial radius will not merge with the primary while a 
  significant fraction of the galaxies found inside of the virial radius 
  will do so by $z\sim0$.

  We note that despite the good agreement between observed and modeled 
  galaxy profiles, the fate of satellite galaxies in numerical simulations 
  should be analyzed with caution.
  In the context of this study, the disagreement between the evolution of
  modeled and observed stellar mass functions 
  (e.g., Fontanot \citeyear{fontanot_many_2009}; 
  Kajisawa \citeyear{kajisawa_moircs_2009};
  Cirasuolo \citeyear{cirasuolo_new_2010};
  Lu \citeyear{lu_bayesian_2012};
  Mutch \citeyear{mutch_constraining_2013})
  suggests that galaxy merging may be misrepresented in numerical calculations.
  Therefore, galaxy merging trajectories in observed group halos may be 
  significantly different from modeled ones.

\section{Summary and conclusions}
 The evolution of galaxy properties is determined in large part by the 
 environments in which they reside and the mass density around them.
 Here we quantified the radial number density profile of 
 galaxies around massive primaries in the redshift range $0.04<z<1.6$.
 We showed that massive galaxies have typically resided in group
 environments in the past 9.5 Gyr and that they are on average surrounded 
 by 2 to 3 massive satellite galaxies.
 We also estimated that the cumulative stellar mass of halo satellites within
 the virial radius accounts for roughly as much mass as in the massive primary 
 itself, suggesting that the potential for mass growth through in-halo mergers 
 is significant.

 Mergers within massive galaxy groups, even at a low rate, are expected to 
 dramatically influence the evolution of observed galaxy distributions in such
 halos, unless an influx of galaxies continuously repopulates the halo.
 We compared the number density profiles of massive satellite galaxies in four 
 redshift bins and showed that the profiles are consistent with no evolution 
 out to $z=1.6$.
 Although this result may be naively interpreted as evidence for the 
 insignificance of galaxy mergers in group halos, the observed mass growth
 of massive quiescent galaxies over the same redshift implies that mergers 
 occur at a non-negligible rate.
 Moreover, results from semi-analytic models suggest that massive primaries
 continuously merge with satellite galaxies in their halo.

 It would also be interesting to compare the observed satellite distributions 
 to predictions from hydrodynamical simulations, such as those of Naab et al.
 (\citeyear{naab_formation_2007}).
 These models successfully reproduce the build-up of the outer envelopes of 
 massive galaxies through minor mergers 
 (e.g., Naab et al. \citeyear{naab_minor_2009}; 
 Oser et al. \citeyear{oser_cosmological_2012};
 Gabor \& Dav\'{e} \citeyear{gabor_growth_2012};
 Hilz et al. \citeyear{hilz_how_2013}), and an important question is whether 
 the reservoir of satellite galaxies is also reproduced.
 
 In conclusion, the observed lack of evolution in the number density profiles
 suggests that there exists a tight balance between mergers and accretion 
 in massive galaxy halos.
 As satellite galaxies merge with their massive primary, other galaxies get
 accreted into the halo at a similar rate but on more extreme trajectories.
 The two competing processes result in a remarkably balanced galaxy 
 distribution out to $r=3000$ kpc, over the redshift range $0<z<1.6$. 
 
 This is the first time that an analysis of the evolution of galaxy
 number density profiles is performed at this redshift and it provides a new
 observational insight on galaxies in group halos.

\begin{acknowledgements}
  We thank Frank van den Bosch, Pascal Oesch and Brad Holden for engaging 
  discussions that contributed to this work.

  This material is based upon work supported by the National Science 
  Foundation under Award No. AST-1202667.

  We gratefully acknowledge support from the CT Space Grant.

  This study makes use of data from the NEWFIRM Medium-Band Survey, a 
  multi-wavelength survey conducted with the NEWFIRM instrument at the 
  KPNO, supported in part by the NSF and NASA.

  Funding for the SDSS and SDSS-II has been provided by the Alfred P. 
  Sloan Foundation, the Participating Institutions, the National Science 
  Foundation, the U.S. Department of Energy, the National Aeronautics and
  Space Administration, the Japanese Monbukagakusho, the Max Planck Society, 
  and the Higher Education Funding Council for England. The SDSS Web Site is
  http://www.sdss.org/.

  The SDSS is managed by the Astrophysical Research Consortium for the 
  Participating Institutions. The Participating Institutions are the 
  American Museum of Natural History, Astrophysical Institute Potsdam, 
  University of Basel, University of Cambridge, Case Western Reserve 
  University, University of Chicago, Drexel University, Fermilab, the 
  Institute for Advanced Study, the Japan Participation Group, Johns 
  Hopkins University, the Joint Institute for Nuclear Astrophysics, the 
  Kavli Institute for Particle Astrophysics and Cosmology, the Korean 
  Scientist Group, the Chinese Academy of Sciences (LAMOST), Los Alamos 
  National Laboratory, the Max-Planck-Institute for Astronomy (MPIA), the 
  Max-Planck-Institute for Astrophysics (MPA), New Mexico State University, 
  Ohio State University, University of Pittsburgh, University of Portsmouth, 
  Princeton University, the United States Naval Observatory, and the 
  University of Washington.

\end{acknowledgements}

  \begin{table*}[t]
    \caption{Projected radial profile measurements}
    \centering
    \begin{tabular}{c c c c c c c c c}
      \hline\hline
       & \multicolumn{2}{c}{$0.04<z<0.07$} & \multicolumn{2}{c}{$0.4<z<0.9$} & 
      \multicolumn{2}{c}{$0.9<z<1.3$} & \multicolumn{2}{c}{$1.3<z<1.6$}\\
      log(R)\footnotemark[1] & log($\phi$)\footnotemark[2] & 
      log($\Delta\phi$)\footnotemark[3] & log($\phi$)\footnotemark[2] &
      log($\Delta\phi$)\footnotemark[3] & log($\phi$)\footnotemark[2] &
      log($\Delta\phi$)\footnotemark[3] & log($\phi$)\footnotemark[2] &
      log($\Delta\phi$)\footnotemark[3]\\
      \hline {\vspace{-5px}}\\
       1.2 &       &       &       &       & -4.14 & -5.18 & -4.10 & -5.22 \\
       1.4 &       &       & -4.35 & -5.20 & -4.35 & -5.26 & -4.31 & -5.24 \\
       1.6 &       &       & -4.26 & -5.42 & -4.70 & -5.57 & -4.79 & -5.44 \\
       1.8 &       &       & -5.01 & -5.51 & -4.89 & -5.69 & -4.78 & -5.69 \\
       2.0 & -5.20 & -5.69 & -4.78 & -5.78 & -4.93 & -5.95 & -5.07 & -5.83 \\
       2.2 & -5.34 & -6.30 & -5.03 & -5.95 & -5.28 & -6.03 & -5.38 & -6.01 \\
       2.4 & -5.60 & -6.32 & -5.29 & -6.08 & -5.73 & -6.34 & -5.45 & -6.25 \\
       2.6 & -5.82 & -6.52 & -5.44 & -6.23 & -5.77 & -6.46 & -5.69 & -6.35 \\
       2.8 & -6.06 & -6.63 & -5.71 & -6.39 & -5.86 & -6.54 & -5.65 & -6.50 \\
       3.0 & -6.26 & -7.09 & -5.94 & -6.40 & -5.98 & -6.67 & -5.78 & -6.55 \\
       3.2 & -6.52 & -6.99 & -6.02 & -6.57 & -6.29 & -6.77 & -5.95 & -6.60 \\
       3.4 & -6.70 & -7.17 & -6.21 & -6.70 & -6.53 & -6.86 & -6.11 & -6.73 \\
       3.6 & -7.09 & -7.37 & -6.40 & -6.83 & -7.15 & -7.01 & -6.38 & -6.79 \\
      \hline\hline
      \label{tab:profiles}
    \end{tabular}
    \footnotetext[1]{Central value of the logarithmic radius bin}
    \footnotetext[2]{Logarithm of the average projected galaxy density 
      in units N$_{\mathrm{gal}}$ kpc$^{-2}$ primary$^{-1}$ dlog(R)}
    \footnotetext[1]{Logarithm of the estimated error of the average projected 
      galaxy density}
  \end{table*}

\bibliographystyle{yahapj}
\bibliography{ms}

\end{document}